\documentclass[aps,twocolumn,prl,superscriptaddress]{revtex4}

\usepackage{graphicx}  
\usepackage{dcolumn}   
\usepackage{bm}        
\usepackage[usenames]{color}
\usepackage[colorlinks,urlcolor=blue,citecolor=blue,linkcolor=blue]{hyperref}
\usepackage{amssymb,amsmath}

\begin{document}

\newtheorem{proposition}{Proposition}
\newtheorem{lemma}{Lemma}
\newtheorem{theorem}{Theorem}
\newtheorem{corollary}{Corollary}
\newtheorem{definition}{Definition}
\newtheorem{remark}{Remark}
\newcommand{\Proof}{\noindent{\bf Proof. }}

\newcommand{\pfreq}{\ensuremath{\Omega}}
\newcommand{\ifreq}{\ensuremath{\Phi}}
\newcommand{\ifreqb}{\ensuremath{\Phi}}
\newcommand{\ifreqbR}{\ensuremath{\Phi_R}}
\newcommand{\rperp}{\ensuremath{R_\perp}}

\newcommand{\rcg}[1]{{\bf{\color{red}$\blacksquare$#1$\blacksquare$}}}
\newcommand{\dsh}[1]{{\bf{\color{blue}$\blacksquare$#1$\blacksquare$}}}

\title{Dynamics of Vortex Dipoles in Confined Bose-Einstein Condensates}

\author{P.\,J.\ Torres}
\affiliation{Departamento de Matem\'atica Aplicada,
Universidad de Granada, 18071 Granada, Spain}
\author{P.\,G.\ Kevrekidis}
\affiliation{Department of Mathematics and Statistics, University of Massachusetts,
Amherst MA 01003-4515, USA}
\author{D.\,J.\ Frantzeskakis}
\affiliation{Department of Physics, University of Athens, Panepistimiopolis,
Zografos, Athens 157 84, Greece}
\author{R.\ Carretero-Gonz\'{a}lez}
\affiliation{
Nonlinear Dynamical System Group,%
\footnote{\tt http://nlds.sdsu.edu}
Computational Science Research Center, and
Department of Mathematics and Statistics,
San Diego State University, San Diego, California 92182-7720, USA}
\author{P.\ Schmelcher}
\affiliation{Zentrum f\"ur Optische Quantentechnologien, Universit\"at
Hamburg, Luruper Chaussee 149, 22761 Hamburg, Germany}
\author{D.\,S.\ Hall}
\affiliation{Department of Physics, Amherst College, Amherst, Massachusetts, 01002--5000 USA}


\begin{abstract}
We present a systematic theoretical analysis of the motion of a pair of straight counter-rotating vortex lines within a trapped Bose-Einstein condensate. We introduce the dynamical equations of motion, identify the associated conserved quantities, and illustrate the integrability of the ensuing dynamics. The system possesses a stationary equilibrium as a special case in a class of exact solutions that consist of rotating guiding-center equilibria about which the vortex lines execute periodic motion; thus, the generic two-vortex motion can be classified as quasi-periodic. We conclude with an analysis of the linear and nonlinear stability of these stationary and rotating equilibria.
\end{abstract}

\maketitle


\section{I. Introduction}
\label{SEC:intro}

Vortices are persistent circulating flow patterns that occur in many diverse scientific and mathematical contexts \cite{Pismen1999}, ranging from hydrodynamics, superfluids, and nonlinear optics to specific instantiations in sunspots, dust devils~\cite{Lugt1983}, and plant propulsion~\cite{Whitaker2010}. In atomic Bose-Einstein condensates (BECs) at
ultracold temperatures \cite{Pethick2002,Pitaevskii2003,Kevrekidis2008}, quantized vortices arise as especially persistent topological defects that play important roles in both Hamiltonian and dissipative dynamics, as well as in quantum turbulence.
Vortices and vortex lattices in atomic BECs
have been analyzed in a series of reviews~\cite{Fetter2001,Kevrekidis2004,Fetter2009}.

Recently, much theoretical~\cite{Crasovan2002,Crasovan2003,Zhou2004,Mottonen2005,Pietila2006,Li2008,Middelkamp2010,Kuopanportti2011} and experimental~\cite{Neely2010,Freilich2010,Seman2010,Middelkamp2011} attention has been devoted to small vortex clusters, with a special focus on the simplest ``vortex molecule,'' the vortex dipole.
This structure, consisting of two countercirculating vortices, is central to the relaxation of superfluids through pairwise reconnection of the vortex lines.
Such reconnections have recently been observed in turbulent superfluid 
helium~\cite{Paoletti2010}, a system considerably more complicated than the dilute-gas BEC. 

Vortices in a Bose-Einstein condensate are observed experimentally by identification of the vortex lines (or ``cores''), which are regions of reduced atomic density that surround the phase singularity about which the fluid rotates. The dynamics of the vortex lines in a dipole are, in general, three-dimensional (3D), and can be simulated through a mean-field model, namely
the Gross-Pitaevskii partial differential equation (PDE)~\cite{Aftalion2003,Aftalion2004,Danaila2005,Neely2010,Rooney2011}. In many cases of experimental interest the vortex lines are straight and parallel, enabling calculations based on two-dimensional (2D) reductions of the PDEs.
Such simulations have not proven entirely satisfactory in their agreement with recent experimental results~\cite{Freilich2010,Middelkamp2011}. An alternative
2D formulation is to focus on the dynamics of the vortex lines themselves by evaluating an ordinary differential equation (ODE) that treats the locations of the lines as individual (quasi) particles~\cite{Middelkamp2010,Middelkamp2011}. The latter approach privileges the vortex lines themselves and enables considerable insight into their fundamental dynamics, even if the loss of the third dimension precludes a direct investigation of reconnection dynamics.

Our aim in the present work is to offer a systematic study of the equations of motion ensuing from the
quasi-particle approach, focusing on an ODE model of two straight counter-circulating vortex lines in a cylindrically symmetric Bose-Einstein condensate. We calculate the conserved quantities within the model and demonstrate its integrability in the case of two vortices. The vortex lines therefore undergo generically quasi-periodic motion. To cement this observation, we present a systematic analysis of the stationary or time-periodic states that the system possesses. We identify time-periodic guiding center solutions and explicitly compute their frequency as well as the fixed distances of their constituent vortex lines from the center of the superfluid. Finally, we examine both the linear and nonlinear stability of such states, arguing that the generic quasi-periodic motion of the two-vortex system consists of such guiding-center rotations along with (second frequency) epicycles around these states.

Our presentation is structured as follows. In Section~II, we present the mathematical model,
while in Section~III, we examine its conserved quantities, integrability and associated dynamics. In Section~IV we consider the special stationary equilibrium, which we generalize to rotating, guiding-center equilibria in Section~V. Finally, we conclude our presentation in Section~VI with some interesting aspects of the problem for future study.

\section{II. Model}

We consider straight line vortices in a Bose-Einstein condensate with cylindrical symmetry. The most prominent physical example arises in the context of harmonically confined, oblate
BECs~\cite{Aftalion2002}, but the same general equations and conclusions may be drawn in the case of a hard-wall container as well~\cite{Fetter2009}. We begin by considering an ensemble of singly-quantized vortices, and in the following section specialize to the case of the vortex dipole.

The motion of a vortex line involves its interactions with other vortex lines, as well as the effect of ``boundary conditions'' associated with the fluid confinement~\cite{Middelkamp2010}. We begin with the interactions. The angular velocity of the fluid flow a distance $r$ from a singly-quantized vortex core in a homogenous fluid is
\begin{align}
\ifreq(r) = \frac{\hbar}{m r^2},
\end{align}
where $m$ is the atomic mass. This flow pattern determines the motion of a second vortex line, which moves with the flow at its center (core)~\cite{Nozieres1999}. The motion of the first vortex line is similarly determined by the flow pattern established by the second. If the separation between the $j$th and $k$th vortex lines is $r_{jk}$, then the two same-charge vortex lines orbit one another at angular frequency $\ifreq(r_{jk})$. On the other hand, if the vortices are counter-circulating (vortices of opposite charge), then the vortex lines move together with linear speed $v_{jk}= r_{jk}\ifreq(r_{jk})$ in the direction of the flow between them (Fig.~\ref{fig:model}a).

\begin{figure}
\centering
\includegraphics[width=0.95\columnwidth]{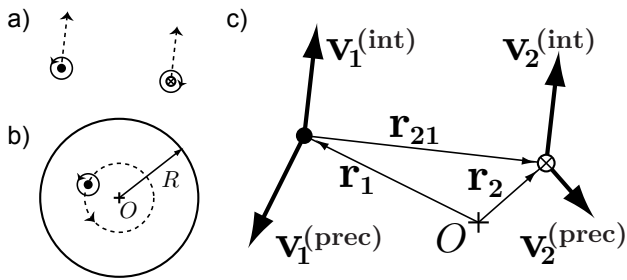}
\caption{\label{fig:model}Elements of the vortex particle model. (a) In an unbounded homogenous fluid, two counter-circulating vortex lines move together (dashed lines) in the direction of the flow between them. (b) A single vortex line in a bounded condensate precesses (dashed line) about the center of the condensate $O$ in the same sense as its circulation. The radius $R$ may be either the Thomas-Fermi radius of the condensate or the wall of the confining cylinder. (c) Two counter-circulating vortex lines in a bounded condensate experience instantaneous velocities $\mathbf{v_{1,2}^{(\text{prec})}}$ and $\mathbf{v_{1,2}^{(\text{int})}}$ associated with precession and interactions, respectively (thick arrows). The thin arrows define the displacements of the two vortices with respect to one another ($\mathbf{r_{21}}$) and the center of the condensate ($\mathbf{r_{1,2}}$).}
\end{figure}

The motion of the $k$th vortex line in either a hard-wall or a disk-shaped harmonic potential also involves gyroscopic precession about the condensate center at fixed distance $r_j$ (Fig.~\ref{fig:model}b). The precession frequency may be expressed in both cases by~\cite{Fetter2001,Fetter2009,Freilich2010}
\begin{align}
\pfreq(r_k) = \frac{\pfreq_0}{1-r_k^2/R^2},
\end{align}
with
\begin{align}
\pfreq_0 = \begin{cases}
\displaystyle\frac{\hbar}{mR^2} &\quad \text{hard-wall,} \\[3.0ex]
\displaystyle\frac{2\hbar\omega_r^2}{8\mu}\left(3 + \frac{\omega_r^2}{5 \omega_z^2}\right)
\ln\left(\frac{2\mu}{\hbar \omega_r}\right) & \quad \text{harmonic,}
\end{cases}
\end{align}
where $\mu$ is the chemical potential, $\omega_r$ and $\omega_z$ are the confining radial and axial frequencies of the harmonic trap, respectively, and $R$ is the Thomas-Fermi radius of the condensate or the radius of the hard-wall cylinder~\cite{Fetter2009}. The common dependence on $r_k$ is what draws together these two different physical situations.

Summing these two velocities for the $k$th vortex in a ``gas" of $n$ vortices gives,
in complex coordinates~\cite{Middelkamp2011},
\begin{equation}\label{eq:ode}
i\dot{z}_k = - S_k \pfreq(r_k) z_k + \frac{b}{2}\sum_{j\neq k}^n S_j \ifreq(r_{jk}) (z_k-z_j),
\end{equation}
where $(x_k,y_k)$ is the position of the $k$th vortex and $z_k = x_k + i y_k=r_k e^{i \theta_k}$, $r_k=|z_k|$, and $r_{jk}=|z_k-z_j|$. The topological charge of the $k$th vortex is $S_k = \pm 1$, with the positive (negative) sign referring to counterclockwise (clockwise) circulation as viewed from the positive $z$ axis. The constant parameter $b$ modifies the interaction strength slightly from the homogeneous case in a harmonic trap; for experimentally relevant parameter values, it has been argued that to a good approximation it is given by $b=1.35$~\cite{Middelkamp2010}, while for the hard-wall potential it is $b=2$.

The relevant velocities and coordinates for a vortex dipole are shown in Fig.~\ref{fig:model}c. Since the model is only valid for straight line vortices, we adopt the convention of referring to the location of a vortex line as the location of the vortex (particle).

\section{III. Conserved Quantities and Integrability}

The system of interest has a Hamiltonian structure. To see this, let us write the differential equations in Cartesian coordinates as
\begin{equation}
\label{eq:ode-cart}
\begin{array}{rcl}
\dot x_k&=&
\displaystyle
- S_k \pfreq(r_k) y_k - \frac{b}{2}\sum_{j\neq k}^n S_j \ifreq(r_{jk}) (y_k-y_j), \\[4.0ex]
\dot y_k&= &
\displaystyle
 S_k \pfreq(r_k) x_k + \frac{b}{2}\sum_{j\neq k}^n S_j
\ifreq(r_{jk}) (x_k-x_j).
\end{array}
\end{equation}
For simplicity we work henceforth in dimensionless distance units with $R=1$. Let us define the function
\begin{eqnarray}
\label{ham}
H(z_1,\ldots,z_n)&=&-\frac{\pfreq_0}{2}\sum_{k=1}^n \ln(1-r_k^2)
\notag
\\[1.0ex]
&&+\frac{\ifreqbR}{4}\sum_{k=1}^n\sum_{j\neq k}^n S_j S_k\ln(r_{jk}^2),
\end{eqnarray}
where $r_k=|z_k|=\sqrt{x_k^2+y_k^2}$ and $\ifreqbR \equiv b\ifreq(R) = \hbar b/m$ with $R=1$.
Then, it can readily be found that
\begin{equation}
\label{orig-ham}
\begin{array}{rcl}
S_k\dot x_k&=&
\displaystyle
-\frac{\partial H}{\partial y_k}, \\[4.0ex]
S_k\dot y_k&=&
\displaystyle
\frac{\partial H}{\partial x_k},
\end{array}
\end{equation}
for every $k=1,\ldots,n$,
which means that $H$ is a first conserved quantity, i.e., the first integral or Hamiltonian along the orbits of the system. Given initial positions $z_k(0)$, then
\begin{align}
H(z_1(t),\ldots,z_n(t))=E_0 \qquad \forall t,
\end{align}
for some suitable constant $E_0$.

It is directly verifiable that a second conserved quantity is
\begin{equation}
\label{momentum}
V=\sum_{k=1}^n S_k r_k^2,
\end{equation}
which represents the angular momentum of the system; see, e.g., Ref.~\cite{Newton2009}
for a discussion about conservation laws (in the absence of the precessional terms).


The
existence of two conserved quantities
guarantees integrability in the classical Liouville
sense~\cite{Arnold1989} for the case $n=2$, whether the vortices are co- or counter-rotating. This implies that the energy level sets are compact and the phase space is foliated by invariant tori. On each of these, the motion is quasi-periodic with two frequencies. In the following, we consider other dynamical aspects of the $n=2$ case for vortex dipoles with $S_1=1$ and $S_2=-1$.


\subsection{A. No Collisions}
The two vortices never collide. To see this, we exploit the fact that the Hamiltonian is constant along orbits. Taking exponentials on the Hamiltonian, we have
\begin{equation}
\label{cons1}
(1-r_1^2)^{\pfreq_0}(1-r_2^2)^{\pfreq_0}r_{12}^{\ifreqbR/2}=C^2>0
\qquad \forall t,
\end{equation}
where $C^2=(1-r_1(0)^2)^{\pfreq_0}(1-r_2(0)^2)^{\pfreq_0}r_{12}(0)^{\ifreqbR/2}$. Note that $0<C^2<2^{\ifreqbR/2}$ because $0\leq r_i(0)^2<1$ ($i=1,2$) and $0<r_{12}(0)<2$. A first consequence of Eq.~(\ref{cons1}) is
\begin{equation}
\label{sep}
r_{12}(t)>C^{4/\ifreqbR} \qquad \forall t,
\end{equation}
that is, the vortices are separated by a computable minimal distance that depends on the initial
position of the vortices.

It is worth reiterating that the model considers the dynamics of straight vortex lines only, such as those occurring in oblate 
BECs, and does not therefore preclude the possibility of collisions and possible reconnection phenomena when the vortex lines are tilted or bent.

\subsection{B. No Ejections}
The vortices never reach the edge of the fluid, remaining confined instead within a
computable inner circle. Assuming $r_{12}(t)<2$ for all $t$ (i.e., the distance
between the two vortices cannot exceed twice the radius, $R=1$, of the condensate),
we have
\begin{equation}
(1-r_1^2)^{\pfreq_0}(1-r_2^2)^{\pfreq_0}=\frac{C^2}{r_{12}^{\ifreqbR/2}}>\frac{C^2}{2^{\ifreqbR/2}}
\qquad \forall t.
\end{equation}
In consequence,
\begin{equation}
r_i^2(t)<1-h,
\end{equation}
where $h=(C^2 2^{-\ifreqbR/2})^{1/\pfreq_0}<1$.

Again, it is worth noting that at finite temperature the dissipative presence of thermal atoms is expected to cause the vortices to leave the fluid. These effects are not considered in the present model.


\section{IV. The Stationary Equilibrium and its Stability}

We now prove the existence of a stationary equilibrium of the vortex pair dynamics and illustrate its stability.

\subsection{A. Existence}

\begin{lemma}
There is an equilibrium, unique up to rotations, given by
\begin{eqnarray}
(x_1^0,y_1^0)&=&\left(\sqrt{\frac{\ifreqbR}{4\pfreq_0+\ifreqbR}},0\right),
\notag
\\[2.0ex]
\notag
(x_2^0,y_2^0)&=&\left(-\sqrt{\frac{\ifreqbR}{4\pfreq_0+\ifreqbR}},0\right).
\end{eqnarray}
\end{lemma}

\Proof  Recall that $n=2$ and $S_1=-S_2=1$. The system (\ref{eq:ode}) in complex notation is
\begin{equation}
\label{orig-complex}
\begin{array}{rcl}
i \dot z_1&=&
\displaystyle
-\pfreq(r_1)\, z_1+ \frac{\ifreqbR}{2 r_{12}^2}(z_1-z_2),
\\[4.0ex]
i \dot z_2&=&
\displaystyle
+\pfreq(r_2)\, z_2- \frac{\ifreqbR}{2 r_{12}^2}(z_2-z_1).
\end{array}
\end{equation}
Here $z_1=r_1\exp(i\theta_1)$ and $z_2=r_2\exp(i\theta_2)$ are, in general, time-dependent, but we look for equilibria, i.e., constant solutions. By direct substitution of the given solution above, one sees that it is indeed a constant solution. Let us prove that this is unique. Naturally, $r_i,\theta_i$ denote polar coordinates (or equivalently an amplitude-phase decomposition).
We can find that the evolution of such polar coordinates reads:
\begin{equation}
\label{orig-polar}
\begin{array}{rcl}
\dot \theta_1&=&\pfreq(r_1)-\displaystyle\frac{\ifreqbR}{2 r_{12}^2}\left[1-\frac{r_2}{r_1}\cos(\theta_1-\theta_2)\right],
\\[4.0ex]
\dot \theta_2&=&-\pfreq(r_2)+\displaystyle\frac{\ifreqbR}{2 r_{12}^2}\left[1-\frac{r_1}{r_2}\cos(\theta_1-\theta_2)\right]
\\[4.0ex]
\dot r_1&=&\displaystyle\frac{\ifreqbR r_2}{2 r_{12}^2}\sin(\theta_1-\theta_2),
\\[4.0ex]
\dot r_2&=&\displaystyle\frac{\ifreqbR r_1}{2
r_{12}^2}\sin(\theta_1-\theta_2).
\end{array}
\end{equation}
Let us call $\theta=\theta_1-\theta_2$. From the last equation, $\sin \theta=0$. Then, $\theta=0$ or $\theta=\pi$. It is straightforward to show that the first option must be discarded. In particular, if $\theta=0$, from the first equation
$$
0=\pfreq(r_1)-\displaystyle\frac{\ifreqbR}{2
r_{12}^2}\left[1-\frac{r_2}{r_1}\right].
$$
Since $\pfreq(r_1)>0$,
$$
1-\frac{r_2}{r_1}=\displaystyle\frac{2 r_{12}^2}{\ifreqbR}
\pfreq(r_1)>0,
$$
and then $r_1>r_2$. Arguing analogously with the second equation, one gets $r_2>r_1$ and thus we reach
a contradiction. In conclusion, it must be the case that $\theta=\pi$. Now, we prove that $r_1=r_2$. On the contrary, let us assume that $r_1>r_2$ (the remaining case is analogous). Adding the two first equations
$$
\pfreq(r_1)-\pfreq(r_2)=\displaystyle\frac{\ifreqbR}{2
r_{12}^2}\left[\frac{r_2}{r_1}-\frac{r_1}{r_2}\right]<0,
$$
which leads to a contradiction because $\pfreq(r)$ is strictly increasing.

In conclusion, $\theta=\pi$, $r_1=r_2$ and we can find the exact value $r_1=r_2=\sqrt{\frac{\ifreqbR}{4\pfreq_0+\ifreqbR}}$ by solving the first equation of 
the system (\ref{orig-polar}).

To emphasize the rotational invariance of our system, it is convenient to work with the variables $(r_1,r_2,\theta)$. The (unique) equilibrium then reads
\begin{equation}
\notag
(r_1^0,r_2^0,\theta_0)=
\left(\sqrt{\frac{\ifreqbR}{4\pfreq_0+\ifreqbR}},\sqrt{\frac{\ifreqbR}{4\pfreq_0+\ifreqbR}},\pi\right).
\end{equation}

\subsection{B. Stability}

\begin{theorem}
The equilibrium $(r_1^0,r_2^0,\theta_0)$ is stable (in the sense of Lyapunov).
\end{theorem}

\Proof
In the coordinates $(r_1,r_2,\theta)$, the system of equations (\ref{orig-complex}) reads
\begin{equation}
\label{orig-polar2}
\begin{array}{rcl}
\dot \theta&=&\pfreq(r_1)+\pfreq(r_2)-\displaystyle\frac{\ifreqbR}{2r_{12}^2}\left[2-\frac{r_1}{r_2}-\frac{r_2}{r_1}\right]\cos\theta,
\\[4.0ex]
\dot r_1&=&\displaystyle\frac{\ifreqbR r_2}{2 r_{12}^2}\sin\theta,
\\[4.0ex]
\dot r_2&=&\displaystyle\frac{\ifreqbR  r_1}{2 r_{12}^2}\sin\theta.
\end{array}
\end{equation}
Of course, $H$ defined by Eq.~(\ref{ham}) is still a conserved quantity; in the new variables, it reads
\begin{eqnarray}
H(r_1,r_2,\theta)&=&-\frac{1}{2}\left[\pfreq_0\ln(1-r_1^2)+\pfreq_0\ln(1-r_2^2)+ \right.
\notag
\\[2.0ex]
\notag
&& \left.
\frac{\ifreqbR}{2}\ln(r_1^2+r_2^2-2r_1r_2\cos\theta)\right].
\end{eqnarray}
Note that basic trigonometrical considerations give $r_{12}^2=r_1^2+r_2^2-2r_1r_2\cos\theta$.

The equilibrium $(r_1^0,r_2^0,\theta_0)$ is a critical point of
$H$. The Hessian matrix evaluated at $(r_1^0,r_2^0,\theta_0)$ is
$$
\frac{1}{8}\left(%
\begin{array}{ccc}
  7\ifreqbR+12\Omega_0+\frac{\ifreqbR^2}{\Omega_0} & \ifreqbR+4\Omega_0 & 0 \\
  \ifreqbR+4\Omega_0 & 7\ifreqbR+12\Omega_0+\frac{\ifreqbR^2}{\Omega_0} & 0 \\
  0 & 0 & \ifreqbR \\
\end{array}%
\right).
$$
One can easily prove that this matrix is positive-definite by Sylvester's criterion. Hence, $H$ is a Lyapunov function that attains a minimum at $(r_1^0,r_2^0,\theta_0)$, and the equilibrium is stable.

The stationary equilibrium is relevant to vortices that are symmetric with respect to a line passing through the condensate center. We illustrate several typical orbits in Fig.~\ref{fig:symmetric}. This class of orbits has been predicted theoretically~\cite{Li2008,Middelkamp2010} and observed experimentally~\cite{Neely2010,Middelkamp2011}.
For relatively small perturbations from the symmetric equilibrium, the (linear) rotational frequency (i.e., the imaginary part of the stability eigenvalue associated with the equilibrium configuration) of these orbits is in reasonable agreement with those found experimentally~\cite{Middelkamp2011}.

\begin{figure}
\centering
\includegraphics[width=0.85\columnwidth]{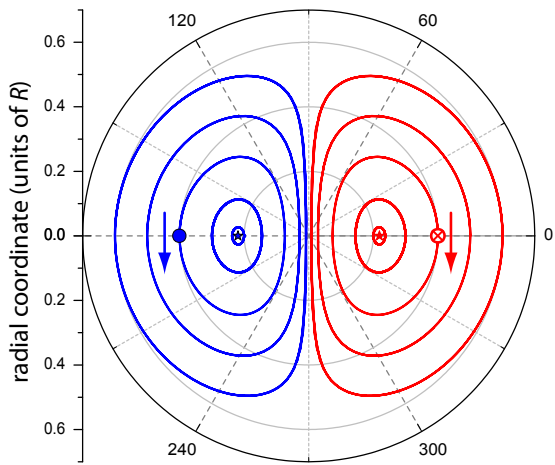}
\caption{
\label{fig:symmetric}
(Color online) Typical computed orbits associated with symmetric vortex configurations, determined by integrating the equations of motion [Eqs.~(\ref{orig-complex})]. The vortex on the left (blue) has a counterclockwise sense of circulation; the vortex on the right (red) has a clockwise sense of circulation. The stationary equilibrium positions of the vortices are indicated by stars. These orbits arise through competition between the effects of interaction and precession~\cite{Li2008}.}
\end{figure}

\section{V. Guiding Center Equilibria and Their Stability}

Finally, we prove the existence of a class of guiding center equilibria and
illustrate their stability.

\subsection{A. Existence}

Consider two vortex positions $(x_1,y_1)$ and $(x_2,y_2)$, or in complex notation, $z_1=r_1\exp(i\theta_1)$ and $z_2=r_2\exp(i\theta_2)$, where all of the $r$'s and $\theta$'s are time-dependent. The system under study is
\begin{align}
i \dot z_1 = -\frac{\pfreq_0}{1-r_1^2} z_1 + \frac{\ifreqbR}{2
r_{12}^2}(z_1-z_2),
\\[2.0ex]
i \dot z_2 = \frac{\pfreq_0}{1-r_2^2} z_2 - \frac{\ifreqbR}{2
r_{12}^2}(z_2-z_1).
\end{align}

We seek precessing guiding centers about which the vortices oscillate. For the special case of guiding centers equidistant from the center of the fluid, we expect the precession frequency $\omega$ to vanish, reproducing the stationary equilibrium case above. Given a vortex positioned at one guiding center, and a (counter-circulating) anti-vortex at the other, we expect that each vortex will stay at its own guiding center; that is, if the guiding centers are themselves precessing, each vortex orbits the center of the condensate at the precession frequency of the guiding centers. From symmetry, we also expect the angular positions of the guiding centers to be $\pi$ radians apart on a line that passes through the center of the condensate, i.e., on opposite sides of the center.

Therefore, we adopt a trial solution of the form
\begin{eqnarray}
z_1(t) &=& r_1 \exp (i\omega t),
\notag
\\[2.0ex]
\notag
z_2(t) &=& r_2 \exp (i\omega t + i\pi) = - r_2\exp(i\omega t),
\end{eqnarray}
where $r_1$ and $r_2$ are now constant, as is $\omega$, the precession frequency of the guiding center. Note that in this case $r_{12}=r_1+r_2 \equiv s$, the constant separation distance between the two guiding centers.

It is now convenient to introduce the notation:
\begin{equation}
\notag
\alpha = \frac{1}{1-r_1^2}, \quad \beta = \frac{1}{1-r_2^2},
\quad \mbox{ and } \quad \gamma = \frac{1}{2 r_{12}^2} =
\frac{1}{2 s^2},
\end{equation}
with $\alpha$, $\beta$, and $\gamma$ all time-independent constants, by assumption.

Since the time dependence in $z_1(t)$ and $z_2(t)$ is now explicit, we take the derivatives directly:
\begin{equation}
\notag
\dot z_1 = i\omega r_1 \exp (i\omega t) \quad \mbox{ and } \quad
\dot z_2 = -i\omega r_2 \exp(i\omega t),
\end{equation}
whereupon the differential equations become
%
%
\begin{eqnarray}
\label{set1}
\omega r_1 &=& \alpha \pfreq_0 r_1 - \gamma \ifreqbR (r_1 + r_2),
\\[2.0ex]
\label{set2}
\omega r_2 &=& -\beta \pfreq_0 r_2 + \gamma \ifreqbR (r_1 + r_2).
\end{eqnarray}
Recalling $s=r_1+r_2=(2\gamma)^{-1/2}$ as a final (as yet unused) constraint, we now have three equations and three unknowns ($r_1$, $r_2$, and $\omega$); that is, upon choosing $s$, the guiding center locations $r_1$ and $r_2$ can be identified
in terms of $s$, as well as the precession frequency of the guiding centers $\omega$ ---provided that there are solutions to this set of equations.
Given the nature of the ensuing algebraic equations (see below), such solutions
will generically exist. We derive the relevant equations below.

We start with the precession frequency $\omega$. From the system of Eqs.~(\ref{set1}) and~(\ref{set2}), we find
%
\begin{align}
\omega = \frac{1}{2}\left[\Omega_0(\alpha-\beta) + \gamma
\ifreqbR\left(\frac{r_1}{r_2}-\frac{r_2}{r_1}\right)\right].
\end{align}
Note that if $r_2=r_1$, then $\alpha=\beta$ and $\omega=0$, recovering the expected stationary equilibrium.

%
Furthermore, from the same Eqs.~(\ref{set1}) and~(\ref{set2}) we immediately obtain:
\begin{align}
\pfreq_0 (\alpha + \beta) - \gamma \ifreqbR \left( 2 + \frac{r_1}{r_2} +
\frac{r_2}{r_1}\right) = 0.
\end{align}
Substituting in for $\alpha$, $\beta$, and $\gamma$, in terms of $r_1$ and $r_2$, and simplifying, one finds
\begin{align}
\pfreq_0 \left(\frac{1}{1-r_1^2}+\frac{1}{1-r_2^2}\right)-\frac{\ifreqbR}{2r_1r_2}=0.
\end{align}
Once the position of $r_2$ is fixed, the location of the first vortex $r \equiv r_1$ can be found from the third order polynomial
\begin{equation}
\beta r^3-\frac{\ifreqbR}{2r_2}r^2- \left(\Omega_0+
\beta\right) r+ \frac{\ifreqbR}{r_2}=0.
\end{equation}
An alternative possibility is to fix $s$, then, the location of the first vortex
$r \equiv r_1$ can be found from the quartic algebraic equation
\begin{equation}
c_4 r^4 + c_3 r^3 +c_2 r^2 + c_1 r + c_0=0,
\end{equation}
where the coefficients $c_j$ ($j=0,1,2,3$) are given by:
\begin{eqnarray}
c_0 &=& B(s^2-1),
\nonumber\\
c_1 &=& 2s-s^3 -2sB ,
\nonumber\\
c_2 &=& 3 s^2 -2 - B(s^2-2),
\\
c_3 &=& -4 s + 2sB,
\nonumber\\
c_4 &=& 2-B, \nonumber \label{sup2}
\end{eqnarray}
and $B\equiv \ifreqbR/(2 \Omega_0)$.

\subsection{B. Stability}

In the following we prove the stability of the guiding center of
the form
\begin{equation}\label{guiding}
z_1(t)=r_1\exp(i\omega t),\qquad z_2(t)=-r_2\exp(i\omega t),
\end{equation}
obtained in the previous subsection. To this purpose, we pass to a
co-rotating frame by making the change of variables $\tilde z_i=z_i\exp(-i\omega
t)$ on the original system \eqref{orig-complex}. Keeping the more
convenient $z_i$ notation for the state variables in a slight
abuse of notation, the resulting system is
\begin{equation}
\label{orig-complex-rot}
\begin{array}{rcl}
i \dot z_1&=&
\displaystyle\omega z_1-\pfreq(r_1)\, z_1+ \frac{\ifreqbR}{2
r_{12}^2}(z_1-z_2), \\[3.0ex]
i \dot z_2&=&
\displaystyle\omega z_2+\pfreq(r_2)\, z_2- \frac{\ifreqbR}{2
r_{12}^2}(z_2-z_1).
\end{array}
\end{equation}
Note that a rotating solution like (\ref{guiding}) of the
original system \eqref{orig-complex} is equivalent to the statement that
$(r_1,-r_2)$ is an equilibrium of the new system (\ref{orig-complex-rot}).

In Cartesian coordinates, the system (\ref{orig-complex-rot}) reads
\begin{equation}
\label{orig-rot}
\begin{array}{rcl}
\dot x_1&=&\omega y_1-\pfreq(r_1)\, y_1+ \ifreqbR\, \displaystyle\frac{y_1-y_2}{2r_{12}^2}, \\[3.0ex]
\dot y_1&=&-\omega x_1+\pfreq(r_1)\, x_1- \ifreqbR\, \displaystyle\frac{x_1-x_2}{2r_{12}^2}, \\[3.0ex]
\dot x_2&=&\omega y_2+\pfreq(r_2)\, y_2- \ifreqbR\, \displaystyle\frac{y_2-y_1}{2r_{12}^2}, \\[3.0ex]
\dot y_2&=&-\omega x_2-\pfreq(r_2)\, x_2+ \ifreqbR\,
\displaystyle\frac{x_2-x_1}{2r_{12}^2},
\end{array}
\end{equation}
and passing to polar coordinates, we obtain
\begin{equation}
\label{orig-polar-rot}
\begin{array}{rcl}
\dot \theta_1&=&-\omega+\pfreq(r_1)-\displaystyle\frac{\ifreqbR}{2r_{12}^2}\left[1-\frac{r_2}{r_1}\cos(\theta_1-\theta_2)\right], \\[4.0ex]
\dot \theta_2&=&-\omega-\pfreq(r_2)+\displaystyle\frac{\ifreqbR}{2r_{12}^2}\left[1-\frac{r_1}{r_2}\cos(\theta_1-\theta_2)\right],\\[4.0ex]
\dot r_1&=&\displaystyle\frac{\ifreqbR r_2}{2r_{12}^2}\sin(\theta_1-\theta_2), \\[3.0ex]
\dot r_2&=&\displaystyle\frac{\ifreqbR
r_1}{2r_{12}^2}\sin(\theta_1-\theta_2).
\end{array}
\end{equation}
Of course, when $\omega=0$ we recover Eqs. \eqref{orig-polar}.

Now the Hamiltonian is
\begin{equation}
\begin{array}{l}
\displaystyle
H(r_1,r_2,\theta)
=-\frac{\omega}{2}(r_1^2-r_2^2)
\\[2.0ex]
\displaystyle
\qquad-\frac{1}{2}
\left[\pfreq_0\ln(1-r_1^2)+\pfreq_0\ln(1-r_2^2)\, + \right.
\\[2.0ex]
\displaystyle
\left.
\qquad\frac{\ifreqbR}{2}\ln(r_1^2+r_2^2-2r_1r_2\cos\theta)\right],
\end{array}
\end{equation}
where again $\theta=\theta_1-\theta_2$. But here the additional difficulty is the absence of a simple explicit expression for the solution.

\begin{lemma}
Any equilibrium of system (\ref{orig-polar-rot}) satisfies $\theta=\pi$.
\end{lemma}

\Proof From the last equation, $\theta=\pi$ or $\theta=0$. By
contradiction, let us assume that $\theta=0$. Then from the two
first equations of system (\ref{orig-polar-rot})
\begin{eqnarray}
-\omega+\pfreq(r_1)-\displaystyle\frac{\ifreqbR}{2r_{12}^2}\left[1-\frac{r_2}{r_1}\right]&=&0,
\label{FirstTwo-orig-polar-rot}
\\[2.0ex]
\notag
-\omega-\pfreq(r_2)+\displaystyle\frac{\ifreqbR}{2r_{12}^2}\left[1-\frac{r_1}{r_2}\right]&=&0,
\end{eqnarray}
subtracting one from the other leads to
$$
\Omega(r_1)+\Omega(r_2)=\frac{\ifreqbR}{2r_{12}^2}\left[2-\frac{r_2}{r_1}-\frac{r_1}{r_2}\right]>0,
$$
then
$$
2-\frac{r_2}{r_1}-\frac{r_1}{r_2}>0
$$
and multiplying by $r_1r_2$ one gets $-(r_1+r_2)^2>0$, which is a
contradiction.

\begin{figure*}
\centering
\includegraphics[width=1.7\columnwidth]{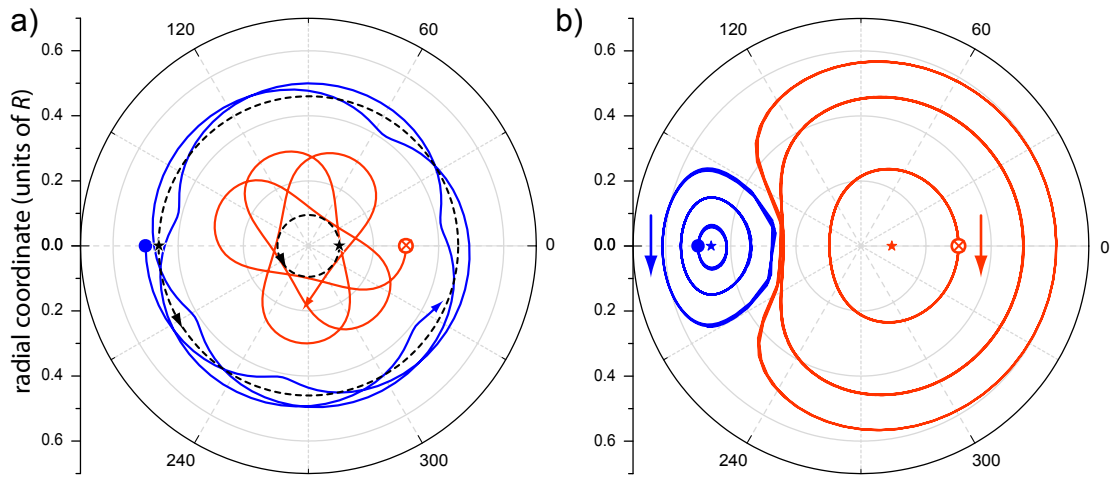}
\caption{
\label{fig:asymmetric}
(Color online) Typical computed orbits associated with asymmetric vortex
configurations, determined by integrating the equations of motion
[Eqs.~(\ref{orig-complex})]. (a) Motion of the vortices in the lab frame,
where the outer vortex (blue) has a counterclockwise sense of circulation
and the inner vortex (red) has a clockwise sense of circulation. The guiding center equilibria are indicated by stars, and their trajectories follow circles in the same sense as the rotation of the outermost vortex (black dashed lines). (b) Motion of vortices in a frame co-rotating with the guiding centers. Several representative orbits with the same guiding centers are shown, with the circle symbols attached to the curves corresponding to the motion in (a).}
\end{figure*}

Now
it is ensured that any equilibrium should be of the form $(r_1,r_2,\pi)$.
We can then compute
the Hessian matrix of the Hamiltonian above, evaluated at the point $(r_1,r_2,\pi)$ and obtain
$$
\left(%
\begin{array}{ccc}
  \pfreq(r_1)+\delta_1 +\delta_{12}-\omega & \delta_{12} & 0 \\
  \delta_{12} & \pfreq(r_2)+\delta_2+\delta_{12}+\omega & 0 \\
  0 & 0 & \delta_{12}r_1 r_2 \\
\end{array}%
\right)
$$
where
$\delta_i\equiv \frac{2 \Omega_0 r_i^2}{(1-r_i^2)^2}>0$ and
$\delta_{12} \equiv \frac{\ifreqbR}{2(r_1+r_2)^2}>0$.
Here it is difficult to see that the matrix is positive-definite,
but, from Eqs.~(\ref{FirstTwo-orig-polar-rot}), the frequency $\omega$ satisfies
\begin{equation}
\notag
\omega=\pfreq(r_1)-\displaystyle\delta_{12}\left(1+\frac{r_2}{r_1}\right)
=-\pfreq(r_2)+\displaystyle\delta_{12}\left(1+\frac{r_2}{r_1}\right).
\end{equation}
Inserting this information into the Hessian yields
$$
\left(%
\begin{array}{ccc}
  \delta_1 +\delta_{12}\left(2+\displaystyle\frac{r_2}{r_1}\right) & \delta_{12} & 0 \\
  \delta_{12} & \delta_2+\delta_{12}\left(2+\displaystyle\frac{r_1}{r_2}\right) & 0 \\
  0 & 0 & \displaystyle\delta_{12} r_1 r_2 \\
\end{array}%
\right)
$$
and now one easily realizes that this matrix is positive-definite again by Sylvester's criterion. Hence, once again the relevant guiding-center equilibrium is a stable one, just as it is in the special case in which $\omega=0$ and $r_1=r_2$.

Rotating guiding center equilibria are relevant when the vortices are asymmetrically located with respect to a line through the condensate center. A representative example is given in Fig.~\ref{fig:asymmetric}; such behavior has been observed experimentally~\cite{Middelkamp2011}.

\section{VI. Conclusions}

From the preceding calculations, we conclude that the vortex dipole system is an integrable one (this is true both in the co-rotating case, and in the counter-rotating one of principal interest herein). This implies that the energy level sets are therefore compact, and the phase space is foliated by invariant tori, on each of which the motion is quasi-periodic with two frequencies. In the context of the vortex dipoles, the first one of these frequencies is the precession of the guiding center equilibria which have been identified herein through the solution of relevant algebraic equations. The positive definite character of the Hessian of the linearization in the appropriate variables ($r_1$, $r_2$ and $\theta=\theta_1-\theta_2$) guarantees stability of these equilibrium points. The second frequency is the oscillation frequency about the precessing equilibria.

Our analysis and conclusions offer a straightforward view of the dynamics of vortex dipoles when the vortex lines are straight, such as those arising in oblate, harmonically trapped Bose-Einstein condensates. Nevertheless,
many interesting questions arise. On one hand, in the counter-rotating vortex case, it seems particularly interesting and relevant to extend the considerations above to the case of three- or more vortex states within the condensate and
investigate the ensuing stationary~\cite{Middelkamp2010} (and perhaps also guiding center) equilibria. On the other hand, another direction that naturally emerges concerns the examination of co-rotating vortices. In the latter case, we certainly expect the generalization of guiding center orbits, due to the common direction of rotation. Understanding the latter phenomenology in the general case of $n$ vortices would arguably be interesting in its own right.


\acknowledgments{This work was supported by the NSF through grants PHY-0855475, DMS-0349023, DMS-0806762, by the M.E.C. of Spain through grant MTM2008-02502 and from the Alexander von Humboldt Foundation.}


\bibliography{dipole}






\end{document}